\documentclass[11pt,a4paper]{article}

\usepackage[utf8]{inputenc} \usepackage{amsmath} \usepackage{amssymb} \usepackage{graphicx} \usepackage[all,dvips]{xy}
\usepackage{amsthm,dsfont} \usepackage[frenchb,english]{babel} \usepackage{ifpdf} \usepackage{color} \usepackage[active]{srcltx}
\usepackage{lmodern} \usepackage{enumerate} \usepackage{fancybox} \usepackage{amsbsy} 
\usepackage{titling} \usepackage{yhmath} \usepackage{lscape} \usepackage{float}  
\usepackage{array} \usepackage{multirow}\usepackage{pdflscape}\usepackage{indentfirst} \usepackage{verbatim}

\usepackage{bibentry}
\nobibliography*
\usepackage{multibib}
\usepackage{multido}
\usepackage{geometry}
\usepackage{caption}
\usepackage{authblk}
\usepackage{titlesec}
\usepackage{enumitem}
\usepackage{marginnote}
\usepackage{setspace} 
\PassOptionsToPackage{hyphens}{url}\usepackage{hyperref}
\hypersetup{
       backref=true,                           
       pagebackref=true,                       
       hyperindex=true,                        
       colorlinks=true,                        
       breaklinks=true,                        
       urlcolor= blue,                         
       linkcolor= blue,                        
       citecolor=blue,			
       bookmarks=true,                         
       bookmarksopen=true,                     
       pdftitle={},  
       pdfauthor={Sebastien Benzekry},                      
       pdfsubject={Mathematics}           
    }

\begin{document}

\title{On the growth and dissemination laws in a mathematical model of metastatic growth}

\author{S\'ebastien Benzekry$^{1}$\footnote{sebastien.benzekry@inria.fr}$\;$ and John ML Ebos$^{2}$}
\date{}

\maketitle
\vskip-1cm
\noindent$^1$inria Bordeaux Sud-Ouest, team MONC and Institut de Math\'ematiques de Bordeaux, Bordeaux, France \\[2ex]
$^2$Departments of Cancer Genetics and Medicine, Roswell Park Cancer Institute, Buffalo, NY, USA

\begin{abstract}
Metastasis represents one of the main clinical challenge in cancer treatment since it is associated with the majority of  deaths. Recent technological advances allow quantification of the dynamics of the process by means of noninvasive techniques such as longitudinal tracking of bioluminescent cells. The metastatic process was simplified here into two essential components -- dissemination and colonization -- which were mathematically formalized in terms of simple quantitative laws. The resulting mathematical model was confronted to \textit{in vivo} experimental data of spontaneous metastasis after primary tumor resection. We discuss how much information can be inferred from confrontation of theories to the data with emphasis on identifiability issues. It is shown that two mutually exclusive assumptions for the secondary growth law (namely same or different from the primary tumor growth law) could fit equally well the data. Similarly, the fractal dimension coefficient in the dissemination law could not be uniquely determined from data on total metastatic burden only. Together, these results delimitate the range of information that can be recovered from fitting data of metastatic growth to already simplified mathematical models.
\end{abstract}

\section{Introduction}

Metastasis (from the greek $\mu\varepsilon\tau\acute{\alpha}$ = change and $\sigma\tau\acute{\alpha}\sigma\iota\zeta$ = place) is a process by which secondary tumors emerge at distant sites from the location of the primary disease \cite{Weiss:2000hl}. This is the result of cells detaching the primary tumor, invading the surrounding tissue, traveling to distant sites (mostly through hematogenous ways) before establishing secondary colonies in distant organs, such as the lungs or the liver (the two organs most often affected nowadays) \cite{Talmadge:2010uva,Weiss:2000hl}. Despite very complex processes happening at various biological scales, the phenomenon can be summarized into two major phases: dissemination and colonization \cite{Chaffer:2011dx,Steeg:2006ih}.

In recent years, novel experimental techniques shed new lights on the process of metastasis \cite{Sahai:2007jb}. On the one hand, the advent of molecular biology revealed the existence and importance of several processes, such as the epithelial-to-mesenchymal transition and its regulation \cite{Chaffer:2011dx,Valastyan:2011vxa,Nguyen:2009vba} or the association between genetic signatures and the probability of metastatic relapse \cite{vandeVijver:2002dt}, of crucial importance considering that metastases are the ones that ultimately kill cancer patients. On the other hand, other processes were discovered that happen at the system scale of the organism. These include the distant suppression of metastatic growth by endogenous inhibitors of angiogenesis \cite{OReilly:1994ik} or the self-seeding phenomenon \cite{Norton:2006kp,Kim:2009tja}. Quantification of the dynamics of these processes is now conceivable thanks to non-invasive experimental techniques that allow to longitudinally track the evolution of the disease \cite{Francia:2011bo}. One of these consist in transfecting cancer cells with an excitable protein such as the luciferin. After orthotopic implantation of these cells in mice, when the enzyme luciferase is subsequently injected to the animals, the cells emit photons, and the emission intensity can be externally recorded.

The advent of these new experimental techniques generates new data that call for quantitative models in order to analyze them and infer general laws of the disease development. Maybe due to the relative scarcity of data about the dynamics of metastasis, relatively few mathematical models have been proposed that could be used to test hypotheses and quantify aspects of the process \cite{Michor:2006hl,Bartoszynski:1987ux,Hanin:2013tha,Retsky:1997vua,Yorke:1993wj,Hartung:2014gz,Haeno:2012fg}. Of note however, in 2000, Iwata, Kawasaki and Shigesada proposed a model particularly well suited for description of the time dynamics of a population of secondary tumors \cite{Iwata:2000gi}. This model can be easily adapted to longitudinal data of total metastatic burden and has been recently shown able to fit such data in several animal models \cite{Hartung:2014gz,Benzekry:2015vz}. Based on one of this study, we will focus here on the identifiability aspects of particular coefficients of the model, respectively related to the growth and dissemination laws. In other words, our aim is to determine what can and what cannot be discriminated based on the data that we dispose on one hand, and the formalism of \cite{Iwata:2000gi} on the other.
\section{Model}
\label{model}
The models that were confronted in this study were largely inspired by the initial model of Iwata, Kawasaki and Shigesada \cite{Iwata:2000gi}. We refer to \cite{Benzekry:2015vz} for a detailed description. Briefly, dynamics of the primary tumor volume $V_p(t)$ at time $t$ is defined by the following Cauchy problem:
\begin{equation}\label{eq:PT}
\left\lbrace 
\begin{array}{l}
\frac{dV_p}{dt} = g_p(V_p(t))\\
V_p(t = 0) = V_i
\end{array}\right.
\end{equation}
where $g_p$ is the primary tumor (PT) growth law (here, either Gompertz or exponential) and $V_i$ is the number of cells at injection, converted into the appropriate unit (either photons per seconds for bioluminescence data or mm$^3$ for caliper-measured volumes). Metastatic development is then defined by two components: the dissemination law $d(V_p)$ and the colonization (or growth) law $g(v)$ for a tumor of volume $v$. These shall be discussed in details below. Specification of these two laws allows to write a partial differential equation for the time evolution of the density of metastases $\rho(t,v)$ structured by the size $v$ of the lesions \cite{Iwata:2000gi}:
\begin{equation}\label{eq:rho}
\left\lbrace\begin{array}{cr}
\partial_t \rho(t,v) + \partial_v(\rho(t,v)g(v)) = 0 & t \in ]0, +\infty[ ,\; v \in ]V_0, +\infty[ \\
g(V_0)\rho(t,V_0) = d(V_p(t)) & t \in ]0, +\infty[\\
\rho(0,v) = 0 & v \in ]V_0, +\infty[ 
\end{array}\right.
\end{equation}
where $V_0$ is the size at which metastasis are born (here assumed to be the size of one cell). For calibration of the conversion ratio from number of cells to bioluminescence, we refer the reader to \cite{Benzekry:2015vz}. In \eqref{eq:rho}, the first equation expresses conservation of the number of metastases when growing in size, the second equates the entering flux of new tumors with the rate of (successful) dissemination from the primary tumor and the last one is the initial condition. From \eqref{eq:rho}, the main quantity of interest for confrontation to bioluminescence data is the total metastatic burden:
$$M(t) = \int_{V_0}^{+\infty}v\rho(t,v) dv = \int_0^t d\left(V_p(t-s)\right)V(s)$$
where $s\mapsto V(s)$ is a solution of the Cauchy problem \eqref{eq:PT} with $g$ instead of $g_p$ and $V_0$ instead of $V_i$. 
\section{Results}

We investigated growth laws being either Gompertz (or Gomp-Exp) or exponential. The dissemination law was assumed to have the following expression (see below for modeling details)
$$d(V_p) = \mu V_p^\gamma$$
and identifiability of $\gamma$ was considered. When not otherwise specified, $\gamma = 1$.
\subsection{Growth law}

\begin{figure}
\begin{center}
\includegraphics[width = 1\textwidth]{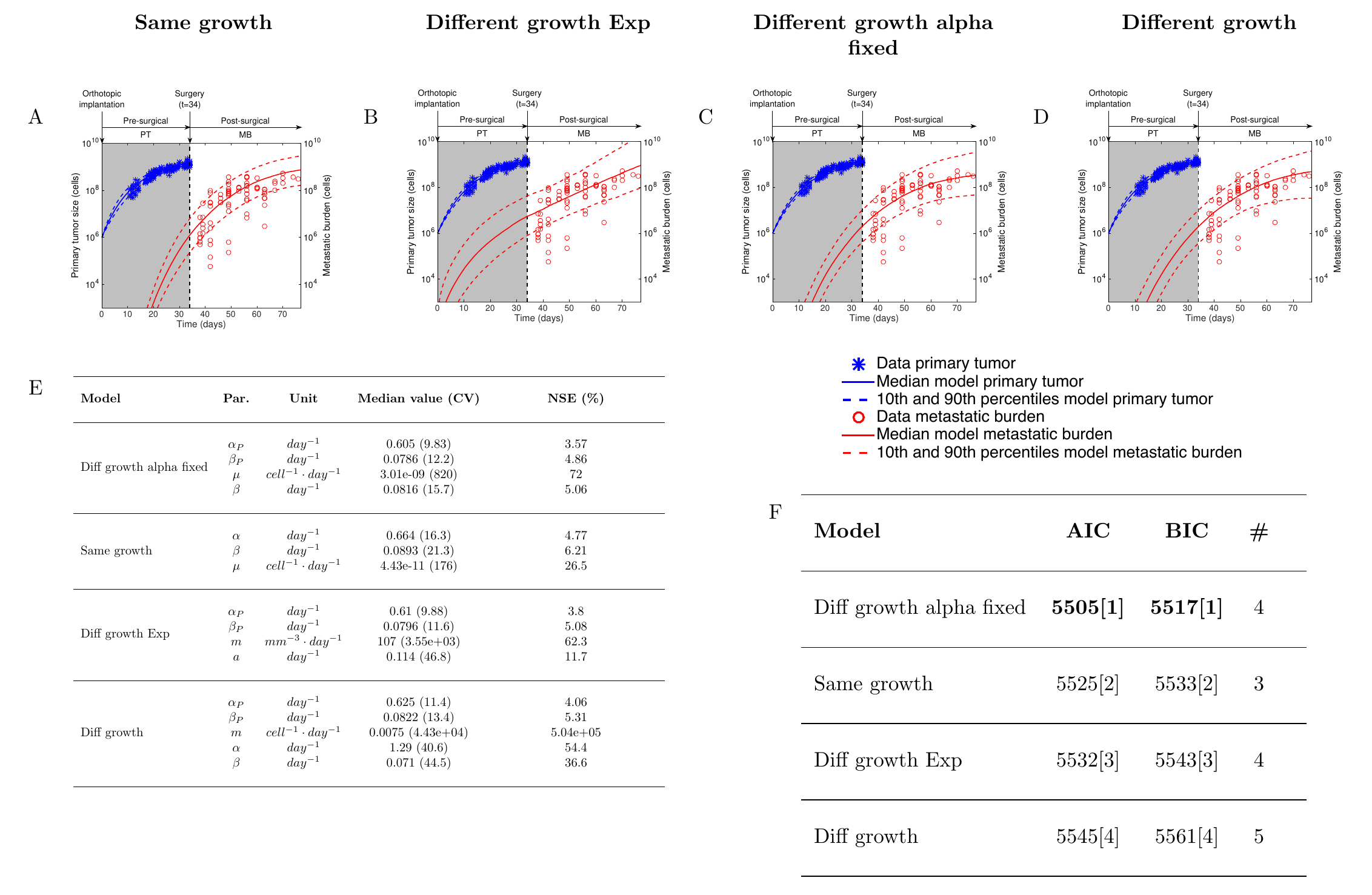}
\end{center}
\caption{
Population fits of the breast xenograft data under different growth theories. 
A: Same growth = same Gompertz growth parameters ($\alpha$ and $\beta$) for primary and secondary tumors.
B: Different growth Exp = exponential growth law for the metastases. 
C: Different growth alpha fixed = for each animal, same value of parameter $\alpha$ was imposed while value of $\beta$ was allowed to vary between the PT and the secondary tumors. 
D: Different growth = the two Gompertz parameters $\alpha$ and $\beta$ were allowed to vary between the PT and the metastases. 
E: Parameters estimates under the two different models, and corresponding normalized standard errors. 
F: Statistical goodness-of-fit metrics for the various models. 
PT = Primary Tumor. MB = Metastatic Burden. CV = coefficient of variation (defined as $\frac{\text{median}}{\text{standard deviation}}\times 100$). NSE = normalized standard error (expressed in percent). AIC = Akaike Information Criterion. BIC = Bayesian Information Criterion. \# = number of parameters.
\label{fig1}
}
\end{figure}

Material and methods generating the data employed here are extensively described in \cite{Benzekry:2015vz}. For the growth of LM2-4$^{\text{luc}+}$ cells orthotopically xenografted in the mammary fat pad of severe combined immune-deficient mice, as employed here, it had been previously demonstrated that the growth kinetics can be accurately described using the Gompertz model \cite{Benzekry:2014kc}, with expression $g_p(v) = v\left( \alpha - \beta \ln\left(\frac{v}{V_0}\right) \right)$. Arguing that for small volumes (such as the one of metastases at initiation), the specific growth rate (i.e. $\frac{g(v)}{v}$) is bounded, and choosing as a reasonable bound the \textit{in vitro} specific growth rate, we rather considered the Gomp-Exp model \cite{Wheldon:1988vua}:
$$g_p(v) = \min\left(\lambda v,  v\left( \alpha - \beta \ln\left(\frac{v}{V_0}\right)\right)\right)$$
where $\lambda$ is the \textit{in vitro} proliferation rate, retrieved from preliminary experiments (see \cite{Benzekry:2015vz}). We then investigated what structural and parametrical shape of the secondary growth coefficient $g(v)$ would better fit the data, while ensuring reasonable identifiability of its parameters. Four scenarii were considered: A) same growth between the PT and the metastases, B) Gomp-Exp for the PT and exponential for the metastases, C) Gomp-Exp for both the PT and the metastases, with $\alpha$ constrained to be identical for the two and D) Gomp-Exp for both the PT and the metastases, with both parameters $\alpha$ and $\beta$ allowed to differ between the PT and the metastases. To use the information at the population level and increase robustness of the parameters estimation (which was found very poor on individual growth curves alone), we used the nonlinear mixed-effects statistical framework to fit the data \cite{Lavielle:2014tw}. Briefly, it consists in maximizing the likelihood of the entire dataset over all individuals, under the assumption of a distribution of the parameters within the population (which was assumed here to be lognormal). Plain and dashed lines in Figure \ref{fig1} are respectively the medians and 10th and 90th percentiles of the outputs of simulations of the model under this distribution.

Selection of the best model was made based on the following considerations: i) visual accuracy of the fit (A-D in Figure \ref{fig1}) and statistical criteria of goodness-of-fit such as the Akaike Information Criterion (F in Figure \ref{fig1}), but also ii) practical identifiability of the parameters, as quantified by the normal standard errors reported in Figure \ref{fig1}.E. As can be observed, model B) has to be rejected for inaccuracy, while model D is clearly over-parameterized (see the values of criteria in Figure \ref{fig1}.F and more importantly the uncertainty on the parameters estimation, especially the normalized standard error on $\mu$ in Figure \ref{fig1}.E). With similar visual accuracy of the population fits (also observable in individual fits of particular mice, results not shown here), the model C) nevertheless generated substantially higher uncertainty on the parameter estimation, especially parameter $\mu$, with respective normalized standard error of 26.5\% for the "same growth" model and 72.1\% for the "different growth with alpha fixed" model. Additionally, probably due to sharper estimation of the parameters, predictive performances were improved with the "same growth" model (results not shown). On the other hand, results of the AIC suggest that the addition of one degree of freedom in model C) yields significant improvement of the goodness-of-fit.

However, under the model C), the difference inferred for parameter $\beta$ translated into minor differences between primary and secondary growth. Together, these results suggest that, although definitive determination of the relationship between the primary and secondary tumor growth law cannot be assessed based only on data on PT and MB growth, the theory assuming equality of the growth rates is conceivable and the most parsimonious.
\subsection{Dissemination law}
With the growth law fixed to the "same growth" model, we investigated determination of the shape of the dissemination law $d(V_p)$. It has been proposed to have the following shape \cite{Iwata:2000gi}:
\begin{equation}\label{eq:diss}
d(V_p) = \mu V_p^\gamma
\end{equation}
with $\mu$ a parameter quantifying the per cell and per day probability of generating a new metastasis, and $\gamma$ possibly related to the fractal dimension of the primary tumor vasculature. Parameter $\gamma$ thus controls the number of cells that are susceptible to leave the tumor while $\mu$ can be linked to a more intrinsic metastatic potential of the disease, aggregating chance of success to multiple steps of the metastatic cascade (including genetic mutations, epithelial-to-mesenchymal transition, blood vessel intravasation, survival in transit, extravasation at the distant site and colonization of the new organ's parenchyma).

\begin{figure}
\begin{center}
\includegraphics[width = 1\textwidth]{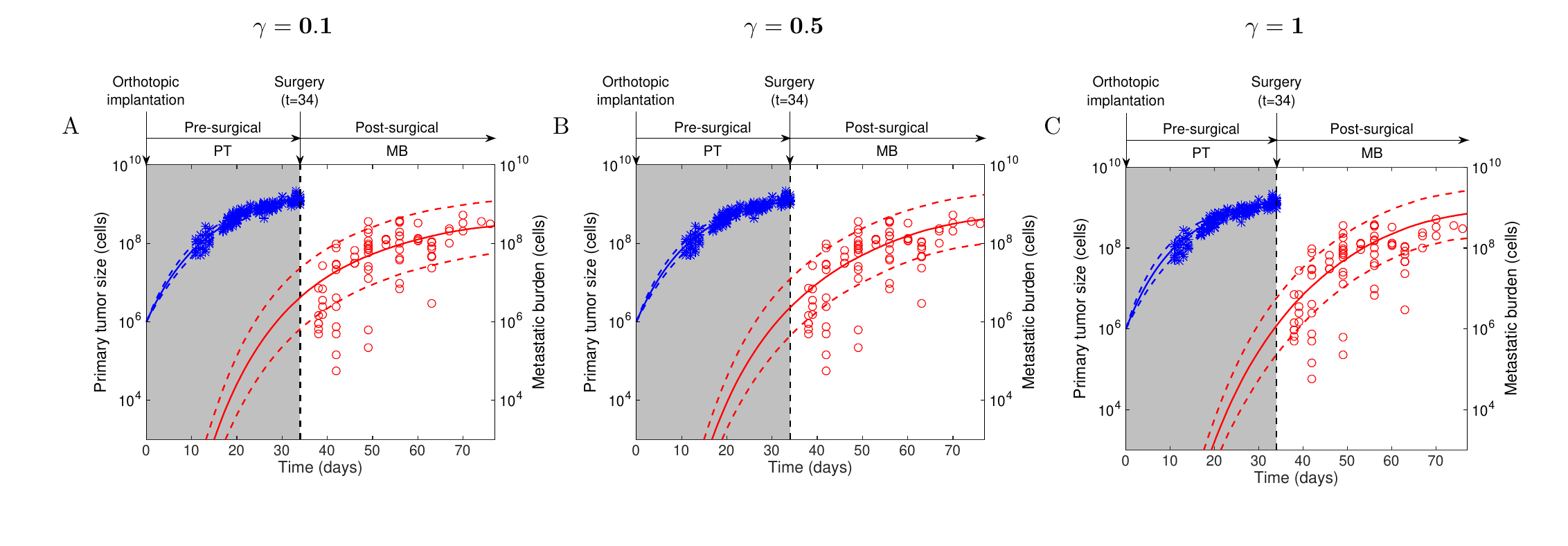}
\end{center}
\caption{\label{fig2}
Population fits of the ortho-surgical metastasis animal models for a dissemination coefficient $\mathbf{d(V_p)=\mu V_p^\gamma}$ and various values of $\mathbf{\gamma}$
}
\end{figure}

We wondered whether the value of $\gamma$ could be determined given the data we dispose. As appears in the results reported in Figure \ref{fig2}, several values generated relatively similar fits. Shown are only three representative values between $0$ and $1$ but every value of $\gamma$ investigated gave similar curves. Although the value of $\gamma = 0.5$ generated a slightly better fit than $\gamma = 0$ or $\gamma = 1$, we did not consider this strong enough to support rejection of either of the possibilities for $\gamma$. We therefore concluded that the precise value of $\gamma$ in \eqref{eq:diss} could not be decided based on our data consisting on PT and post-surgery metastatic growth.

\section{Discussion}

Any data-based modeling inference is limited by two major aspects: 1) the richness of the data and 2) the complexity of the model. Only when taking into account the balance between these two aspects can we reject hypotheses and thus improve our  knowledge about natural mechanisms.

Metastasis is a complex process that remains poorly understood, especially during the colonization phase (after extravasation at the distant site). Using a mathematical formalism of basic laws of the process (dissemination and growth), we evaluated here the limits of what can be learned from longitudinal noninvasive data of post-surgery spontaneous metastasis. Our results demonstrated that similar growth between the primary tumor and the metastases was a conceivable theory for metastatic development in a xenograft breast cancer animal model. Exponential growth of the metastases had to be rejected while, on the other hand, a model with different growth of the metastases could also explain the data. Further quantification of metastatic growth are required to discriminate between the "same growth" and "different growth alpha fixed" models, although growth of secondary tumors was not substantially different when inferred under the second model. To this end, longitudinal follow up of individual metastatic lesions in mice using magnetic resonance imaging might be of great help \cite{Baratchart:2015wr}. Non-uniqueness was also observed for the values of the fractal dimension of the vasculature in the dissemination coefficient able to fit the data. This justifies the use of $\gamma = 1$, i.e. the simplest theory (all cells within the primary tumor have equal probability of generating a metastasis) to describe the data \cite{Benzekry:2015vz}. Additional data are required to discriminate if a particular value of $\gamma$ is better adapted for specification of the dissemination law.

Together, our results demonstrate that, even with a simple mathematical framework (3 degrees of freedom in the most adapted version) and relatively rich data, several mutually exclusive models can fit data of total metastatic burden dynamics, leaving open the question of definitive determination of the growth and dissemination laws of metastatic development.
\bibliographystyle{apalike}
\bibliography{LibraryPapers}
\end{document}